\documentclass[10pt,leqno]{amsart}
\usepackage{graphicx}
\usepackage{indentfirst,csquotes}

\usepackage{amssymb,amsthm,amsmath}
\usepackage{xcolor,paralist,hyperref,fancyhdr,etoolbox}
\usepackage{booktabs}
\usepackage{multirow}
\usepackage{tikz}
\usetikzlibrary{positioning}
\usepackage{pgfplots}
\pgfplotsset{compat=1.17}
\usepackage[linesnumbered,ruled,vlined]{algorithm2e}

\hypersetup{ colorlinks=true, linkcolor=black, filecolor=black, urlcolor=black }

\makeatletter
\patchcmd{\@settitle}{\uppercasenonmath\@title}{}{}{}
\patchcmd{\@setauthors}{\MakeUppercase}{}{}{}
\patchcmd{\@setauthors}{\scshape}{}{}{}
\makeatother

\begin{document}
\title{Hybrid Classical-Quantum Transfer Learning with Noisy Quantum Circuits}

\author[D. Martín-Pérez et al.]{
    \normalfont
    D. Martín-Pérez$^{1}$, 
    F. Rodríguez-Díaz$^{1}$, \\
    D. Gutiérrez-Avilés$^{2}$, 
    A. Troncoso$^{1}$, 
    F. Martínez-Álvarez$^{1}$ \\[1ex]
     \texttt{ $^{1}$ Data Science and Big Data Lab, Pablo de Olavide University, Seville, Spain \\
     $^{2}$ Department of Computer Science, University of Seville, Seville, Spain} \\[1ex]
     \texttt{\{dmarper2, froddia, atrolor, fmaralv\}@upo.es, dgutierrez3@us.es}
}

\date{\today}

\begin{abstract}
    Quantum transfer learning combines pretrained classical deep learning models with quantum circuits to reuse expressive feature representations while limiting the number of trainable parameters. In this work, we introduce a family of compact quantum transfer learning architectures that attach variational quantum classifiers to frozen convolutional backbones for image classification. We instantiate and evaluate several classical-quantum hybrid models implemented in PennyLane and Qiskit, and systematically compare them with a classical transfer-learning baseline across heterogeneous image datasets. To ensure a realistic assessment, we evaluate all approaches under both ideal simulation and noisy emulation using noise models calibrated from IBM quantum hardware specifications, as well as on real IBM quantum hardware. Experimental results show that the proposed quantum transfer learning architectures achieve competitive and, in several cases, superior accuracy while consistently reducing training time and energy consumption relative to the classical baseline. Among the evaluated approaches, PennyLane-based implementations provide the most favorable trade-off between accuracy and computational efficiency, suggesting that hybrid quantum transfer learning can offer practical benefits in realistic NISQ era settings when feature extraction remains classical.    

\end{abstract}

\maketitle
\section{Introduction}
The rapid advancement of artificial intelligence (AI) and machine learning (ML) over the past decade has significantly transformed numerous sectors, including healthcare, industry, finance, and environmental management. One of the key elements behind these advancements has been deep learning \cite{LECUN15}, particularly convolutional neural networks (CNNs), which have demonstrated exceptional capabilities in image recognition, natural language processing (NLP), and complex decision-making tasks. Despite their success, traditional deep learning models often suffer from critical limitations, such as the need of large annotated datasets, high computational demands, and substantial energy consumption during training and deployment phases \cite{SCHWARTZ20}. These constraints have motivated the exploration of alternative learning paradigms that can improve efficiency and sustainability without sacrificing predictive performance.

Transfer learning has emerged as an effective strategy to address some of these limitations. By reusing knowledge learned from models trained on large-scale datasets, transfer learning reduces both the data requirements and computational cost of adapting models to new tasks \cite{Pan2010}. Classical–classical (CC) transfer learning, in which pretrained neural networks are fine-tuned on smaller target datasets, has therefore become a standard approach across a wide range of applications. However, even in this setting, fine-tuning large architectures with millions of parameters remains computationally expensive and energy-intensive, particularly when applied repeatedly across tasks \cite{DWIVEDI25}.

Parallel to these developments, quantum machine learning (QML) \cite{RODRIGUEZ26} has attracted increasing attention as a potential paradigm for enhancing machine learning workflows. By exploiting quantum-mechanical phenomena such as superposition and entanglement, quantum computing promises computational advantages for specific problem classes. At present, however, quantum hardware operates in the noisy intermediate-scale quantum (NISQ)~\cite{Preskill2018} regime, characterized by limited qubit counts, short coherence times, and restricted circuit depths. These constraints significantly limit the direct applicability of fully quantum algorithms to real-world machine learning problems \cite{BHARTI22}.

To overcome these limitations, hybrid classical–quantum  (CQ) approaches have been proposed. Within this context, Quantum Transfer Learning (QTL), and in particular CQ transfer learning, has emerged as a promising direction. In QTL, pretrained classical neural networks act as fixed feature extractors, while compact quantum circuits are employed as trainable classification or regression heads. This architectural separation enables quantum models to operate on low-dimensional representations that retain the most informative features, improving parameter efficiency and mitigating current hardware constraints.

QTL remains comparatively underexplored from a systematic benchmarking perspective, particularly with respect to direct comparisons against well-established CC transfer-learning baselines under realistic experimental conditions. 
In this work, we address this gap by conducting a controlled, systematic evaluation of CC and CQ transfer learning pipelines, assessing their performance, efficiency, and robustness across heterogeneous image classification tasks from medical, biological, industrial, and general-vision domains.

Our analysis exploits widely used pretrained CNN backbones and focuses on isolating the impact of the quantum component, comparing predictive accuracy, training time, and energy consumption across classical and hybrid models. In addition, we explicitly account for realistic quantum noise by incorporating hardware-calibrated noise models, thereby enabling a more accurate assessment of near-term quantum machine learning performance. 

In summary, our work makes the following contributions to QTL: 
\begin{enumerate}
\item We introduce a family of new QTL architectures that take features from pretrained convolutional backbones and feed them into compact variational quantum classifiers, instantiated in both PennyLane and Qiskit.

\item We design and apply a fair evaluation protocol that addresses classical feature extractors, spans heterogeneous image datasets, and incorporates realistic noise models calibrated on IBM hardware to compare ideal simulation with noisy emulation.

\item We present a head-to-head comparison of several quantum-enhanced variants and a classical transfer learning baseline, reporting accuracy across multiple backbones and datasets.


\item We analyze the effect of realistic noise on model performance and identify the need for error mitigation strategies in near-term devices.

\item We release implementations and experimental configurations to enable reproducible benchmarking of hybrid CQ systems.
\end{enumerate}

The remainder of the paper is structured as follows. Section \ref{sec:Related} reviews related work on transfer learning and hybrid quantum classifiers. Section \ref{sec:methodology} presents our methodology, detailing the proposed QTL architectures in PennyLane and Qiskit. Section \ref{sec:results} presents the experimental results, comparing the quantum variants with the classical baseline under ideal simulation and realistic noise conditions. Finally, Section \ref{sec:conclusions} concludes with a discussion of implications and future directions.

\section{Related Work}\label{sec:Related}
In this section, we review prior work on QTL, with particular emphasis on hybrid CQ architectures, application domains, and experimental evaluation practices under NISQ constraints.

A first line of research explores QTL for image-based classification tasks, particularly in medical and scientific domains. Bali et al. \cite{Bali25} introduce QuantumNet, a hybrid framework combining classical feature extractors with variational quantum classifiers for diabetic retinopathy detection. Mir et al. \cite{Mir21} similarly combine pretrained convolutional backbones with variational quantum classifiers, evaluating their approach across multiple quantum software frameworks. Mogalapalli et al. \cite{Mogalapalli21} study QTL for diverse image classification tasks, showing that the suitability of classical backbones depends strongly on dataset characteristics and training configurations. Azevedo et al. \cite{Azevedo22} further validate the applicability of QTL in medical imaging through experiments on real quantum hardware, demonstrating robustness beyond ideal simulation.

Beyond medical imaging, QTL has also been applied to general vision and industrial datasets. Kumsetty et al. \cite{Kumsetty22} introduce the TrashBox dataset and evaluate hybrid CQ transfer learning models for waste classification, reporting improvements in both predictive performance and training efficiency. Otgonbaatar et al. \cite{Otgonbaatar23} address the challenges of high-dimensional embeddings and limited qubit availability by combining pretrained VGG16 features with multiqubit quantum layers, showing improved generalization through strongly entangling circuits. Kati et al. \cite{Kati25} extend QTL to deepfake detection by integrating it with transformer-based vision architectures, demonstrating the compatibility of hybrid quantum models with modern attention-driven feature extractors.

A complementary body of work investigates QTL across non-visual modalities, highlighting its applicability beyond image classification. Thus, Qi et al. \cite{Qi22} propose a CQ transfer learning approach for spoken command recognition using pretrained audio representations. Wang et al. \cite{Wang23} apply QTL to synthetic speech detection by fine-tuning variational quantum circuits on top of large pretrained speech models. Koike et al. \cite{Koike22} extend QTL to wireless sensing to mitigate domain shift in human activity monitoring using 60-GHz Wi-Fi signals. In the NLP domain, Buonaiuto et al. \cite{Buonaiuto24} explore hybrid CQ classifiers for linguistic acceptability judgments, achieving results comparable to state-of-the-art classical models.

QTL has also been explored in sequential and time-series modeling tasks. Wang et al. \cite{Wang24} propose a layer-enhanced quantum LSTM combined with transfer learning for remaining useful life prediction in fuel cells, demonstrating improved predictive stability under limited data regimes. In a related but distinct setting, Han et al. \cite{Han23} introduce a quantum-inspired transfer learning approach for control optimization in wind turbines, achieving substantial runtime reductions compared to conventional optimization techniques.

Several works focus on architectural strategies to adapt quantum models to NISQ-era constraints. Kim et al. \cite{Kim23} integrate pretrained CNNs with quantum convolutional neural networks to reduce circuit depth while maintaining classification accuracy. Yogaraj et al. \cite{Yogaraj25} propose post-variational classical QTL to alleviate optimization challenges in variational quantum circuits, improving robustness across multiple backbones and datasets. Khatun et al. \cite{Khatun25} further investigate QTL architectures with an emphasis on adversarial robustness, demonstrating improved resilience compared to both classical and fully quantum alternatives.

From a broader perspective, several studies explore transfer learning in quantum or quantum-inspired settings beyond standard supervised classification. Zen et al. \cite{Zen20} use transfer learning to improve the scalability of neural-network quantum states in many-body physics, while Vermeire et al. \cite{Vermeire21} apply quantum-informed transfer learning to chemical property prediction by combining quantum-calculated and experimental data, achieving improved accuracy and generalization in data-limited regimes.

Despite these advances, existing QTL studies typically evaluate a limited number of architectures or datasets, often rely on idealized quantum simulations, and rarely report computational cost or energy consumption as first-class evaluation metrics. Moreover, comparisons across different quantum software frameworks are frequently conducted in isolation, making it difficult to draw general conclusions about practical performance trade-offs in realistic NISQ settings.

In contrast, our work provides a controlled and systematic benchmarking of classical–classical and CQ transfer learning pipelines across multiple datasets, pretrained backbones, quantum architectures, and software frameworks. By incorporating realistic, hardware-calibrated noise models and explicitly measuring training time and energy consumption, we aim to assess QTL not only in terms of predictive accuracy but also from the perspective of practical deployment and sustainable computing.

\section{Methodology}
\label{sec:methodology}
This work presents a comparative analysis of four distinct image classification architectures, obtained by combining multiple pretrained classical backbones with three families of transfer-learning heads, including a classical baseline and two quantum-hybrid variants. Our objective is not to demonstrate quantum advantage, but to assess whether compact quantum heads can act as effective and efficient alternatives to classical classifiers under realistic NISQ constraints. To this end, we evaluate the impact of realistic quantum noise on hybrid model performance through controlled experiments, ensuring fair comparison across all architectures.

Classical baselines employ pretrained convolutional networks as frozen backbones, with only their final classifier layers fine-tuned using cross-entropy loss and ImageNet normalization. These models establish the reference accuracy and energy profile without any quantum layer.

Qiskit-based quantum hybrids feed the same backbones into a \texttt{SamplerQNN} constructed with four qubits and a depth-three variational circuit, employing Hadamard initialization, angle encoding via $R_Y$ gates, and brick-wall CNOT entanglement patterns alternating between even-odd qubit pairs~\cite{Mitarai2018}. Each variational layer applies parameterized $R_Y$ rotations followed by entangling operations, with outputs interpreted through parity-based binary string classification. Optional shot sampling ($N_{\text{shots}}=1024$) and depolarizing noise (single-qubit: $p_{1q}=0.001$, two-qubit: $p_{2q}=0.01$) emulate realistic NISQ hardware~\cite{Preskill2018}. 

PennyLane-based quantum hybrids project backbone features to a four-qubit quantum layer implemented with \texttt{AngleEmbedding} followed by \texttt{BasicEntanglerLayers} using circular CNOT topology~\cite{Bergholm2018pennylane}. The circuit consists of three variational layers, each containing circular entanglement (qubits $0 \to 1 \to 2 \to 3 \to 0$) and parameterized $R_Y$ rotations, with Pauli-$Z$ expectation values on all qubits replacing the classical fully connected head. The noisy variant applies realistic IBM noise channels (amplitude and phase damping) after each gate operation using the \texttt{default.mixed} device with parameters calibrated from IBM Heron r2 specifications (T1=250$\mu$s, T2=150$\mu$s). Across all scenarios, we freeze the convolutional parameters, standardize optimizer settings (Adam with lr=$10^{-3}$, StepLR scheduler with $\gamma=0.9$), and log losses, accuracies, and confusion matrices to enable direct comparisons of classical and quantum-enhanced transfer learning.

\subsection{Classical-Classical Transfer Learning}
\label{subsec:classical_tl}
Our classical baseline employs transfer learning~\cite{Pan2010,Yosinski2014,Zhuang2021} using CNNs pretrained on ImageNet-1k~\cite{Deng2009,Russakovsky2015}. We evaluate several pretrained architectures that vary in design philosophy, parameter efficiency, and feature extraction capabilities~\cite{Goodfellow2016}.

In the training protocol, all backbone weights remain frozen, whereas only the final classification layer is optimized~\cite {Yosinski2014}. The classifier implements a simple linear transformation that maps extracted features to class logits, as defined in Eq. (\ref{eq:classical}).

\begin{equation}
    f_{\text{class}}(\mathbf{x}) = \mathbf{W}^\top \phi(\mathbf{x}) + \mathbf{b}
\label{eq:classical}
\end{equation}

where $\phi(\mathbf{x})$ represents the frozen backbone features, and $\mathbf{W} \in \mathbb{R}^{d \times C}$ defines the trainable weight matrix, with $d$ being the feature dimension and $C$ the number of classes.

\begin{algorithm}
\SetAlgoLined
\KwIn{Dataset $\mathcal{D}$, Pretrained Model $\mathcal{M}_{pre}$, Classes $C$}
\KwOut{Trained Classical Model $\mathcal{M}$}
Initialize $\mathcal{M} \leftarrow \mathcal{M}_{pre}$ (e.g., ResNet18)\;
\ForEach{parameter $p \in \mathcal{M}_{backbone}$}{
    $p.requires\_grad \leftarrow \text{False}$ \tcp*{Freeze backbone}
}
$d_{in} \leftarrow \text{GetInputFeatures}(\mathcal{M}_{classifier})$\;
$\mathcal{M}_{classifier} \leftarrow \text{Linear}(d_{in}, C)$ \tcp*{Replace head}
\While{epoch $\le$ max\_epochs}{
    \ForEach{batch $(x, y) \in \mathcal{D}_{train}$}{
        $f \leftarrow \mathcal{M}_{backbone}(x)$ \tcp*{Extract features}
        $logits \leftarrow \mathcal{M}_{classifier}(f)$\;
        $\mathcal{L} \leftarrow \text{CrossEntropy}(logits, y)$\;
        Update $\mathcal{M}_{classifier}$ via Gradient Descent\;
    }
}
\caption{Classical Transfer Learning Training.}
\label{alg:classical_tl}
\end{algorithm}

\subsection{Classical-Quantum PennyLane (Ideal)}
\label{subsec:pennylane_standard}

The PennyLane Standard architecture integrates a parameterized quantum circuit as the classification layer while retaining a frozen classical backbone for feature extraction~\cite{Mari2020,Mogalapalli21}. The hybrid pipeline transforms input images through the following sequence Eq. (\ref{eq:pennylaneideal}).

\begin{equation}
    \mathbf{x} \xrightarrow{\phi_{\text{CNN}}} \mathbf{f} \xrightarrow{\mathbf{W}_{\text{pre}}} \mathbf{z} \xrightarrow{\tanh(\cdot) \times \frac{\pi}{2}} \boldsymbol{\theta} \xrightarrow{\text{QC}} \langle \hat{O}_i \rangle \xrightarrow{\mathbf{W}_{\text{post}}} \mathbf{y}
\label{eq:pennylaneideal}
\end{equation}

Classical features $\mathbf{f} \in \mathbb{R}^d$ from the frozen backbone undergo linear projection via $\mathbf{W}_{\text{pre}} \in \mathbb{R}^{n \times d}$ (d is the depth of the circuit) to match the quantum layer dimensionality of $n=4$ qubits. The hyperbolic tangent activation scaled by $\pi/2$ ensures angle-encoded inputs $\boldsymbol{\theta} \in [-\pi/2, \pi/2]^n$ remain within valid rotation ranges. The parameterized quantum circuit $\text{QC}(\boldsymbol{\theta}, \boldsymbol{\phi})$ processes these encoded features, producing expectation values $\langle \hat{O}_i \rangle$ of Pauli-$Z$ observables on all qubits. Finally, a post-quantum linear layer $\mathbf{W}_{\text{post}} \in \mathbb{R}^{C \times n}$ maps these quantum measurements to class logits $\mathbf{y}$.

We employ a variational quantum circuit with $n=4$ qubits and depth $D=3$~\cite{Mitarai2018,Havlicek2019}. We select those numbers because the noise is lower than with more qubits, achieving the same or higher accuracy. We tested 1 to 16 qubits, and 4 qubits were the best option. Implemented using PennyLane's \texttt{AngleEmbedding} and \texttt{BasicEntanglerLayers} templates. Classical features are first encoded into quantum states via single-qubit $R_Y$ rotations~\cite{Schuld2015}, mapping input data $\boldsymbol{\theta}$ into quantum-state amplitudes via Eq. (\ref{eq:angleembedding}).

\begin{equation}
    \hat{U}_{\text{enc}}(\boldsymbol{\theta}) = \prod_{i=0}^{n-1} R_Y^{(i)}(\theta_i)
\label{eq:angleembedding}
\end{equation}

The subsequent variational layers combine entanglement and parameterized transformations. Each of the $D=3$ layers implements a circular CNOT pattern, creating ring-topology connectivity via Eq. (\ref{eq:entanglement}).

\begin{equation}
    \hat{U}_{\text{ent}} = \text{CNOT}_{0,1} \cdot \text{CNOT}_{1,2} \cdot \text{CNOT}_{2,3} \cdot \text{CNOT}_{3,0}
\label{eq:entanglement}
\end{equation}

This circular arrangement ensures maximal qubit connectivity~\cite{Sim2019}, where the last qubit wraps around to entangle with the first, forming a complete ring. Following entanglement, trainable $R_Y$ gates apply parameterized rotations~\cite{Farhi2018} using Eq. (\ref{eq:circulararrangement}). 

\begin{equation}
    \hat{U}_{\text{rot}}^{(l)}(\boldsymbol{\phi}_l) = \prod_{i=0}^{n-1} R_Y^{(i)}(\phi_{l,i})
\label{eq:circulararrangement}
\end{equation}

where $\boldsymbol{\phi}_l = (\phi_{l,0}, \phi_{l,1}, \phi_{l,2}, \phi_{l,3})$ represents trainable parameters for layer $l$. The complete circuit combines these operations sequentially, as shown in Eq. (\ref{eq:pennylanecircuit}).

\begin{equation}
    \hat{U}_{\text{circuit}}(\boldsymbol{\theta}, \{\boldsymbol{\phi}_l\}) = \prod_{l=0}^{D-1} \left[\hat{U}_{\text{rot}}^{(l)}(\boldsymbol{\phi}_l) \cdot \hat{U}_{\text{ent}}\right] \cdot \hat{U}_{\text{enc}}(\boldsymbol{\theta})
\label{eq:pennylanecircuit}
\end{equation}

Measurement extracts outputs as expectation values of the Pauli-$Z$ operator on all qubits using Eq. (\ref{eq:pennylaneZ}).

\begin{equation}
    \langle \hat{O}_i \rangle = \langle \psi | \hat{Z}_i | \psi \rangle, \quad i = 0, 1, 2, 3
\label{eq:pennylaneZ}
\end{equation}

yielding a 4-dimensional feature vector for the post-quantum classifier using Pauli Z. The total number of trainable quantum parameters amounts to $n \times D = 4 \times 3 = 12$. Circuit simulation employs PennyLane's \texttt{default.qubit} device~\cite{Bergholm2018pennylane}, performing noiseless statevector evolution. Gradient computation utilizes the parameter-shift rule~\cite{Mitarai2018} using Eq.  (\ref{eq:pennylanegradient}).

\begin{equation}
    \frac{\partial \langle \hat{O} \rangle}{\partial \phi_i} = \frac{1}{2}\left[\langle \hat{O} \rangle_{\phi_i + \pi/2} - \langle \hat{O} \rangle_{\phi_i - \pi/2}\right]
\label{eq:pennylanegradient}
\end{equation}

This analytical approach enables exact gradient evaluation, facilitating efficient variational parameter optimization through standard backpropagation.

\begin{algorithm}
\SetAlgoLined
\KwIn{Features $f$, Qubits $n$, Layers $D$}
\KwOut{Expectation Values $\langle Z \rangle$}
Initialize Device `default.qubit`\;
Define Circuit $U(\theta, \phi)$:\;
\Indp
    $\text{AngleEmbedding}(\theta)$ \tcp*{Encode data}
    \For{$l \gets 1$ \KwTo $D$}{
        $\text{BasicEntanglerLayers}(\phi_l)$ \tcp*{Circular CNOTs + Rot}
    }
    \Return $[\langle Z_0 \rangle, \dots, \langle Z_{n-1} \rangle]$\;
\Indm
\textbf{Forward Pass:}\;
\Indp
    $x_{proj} \leftarrow W_{pre} \cdot f$ \tcp*{Reduce dimension}
    $\theta \leftarrow \tanh(x_{proj}) \cdot \frac{\pi}{2}$ \tcp*{Scale to $[-\pi/2, \pi/2]$}
    $q_{out} \leftarrow \text{Execute } U(\theta, \phi)$\;
    $logits \leftarrow W_{post} \cdot q_{out}$\;
\Indm
\caption{Classical-Quantum PennyLane (Ideal).}
\label{alg:pl_standard}
\end{algorithm}

\begin{figure}[htbp]
    \centering
    \includegraphics[width=0.7\textwidth]{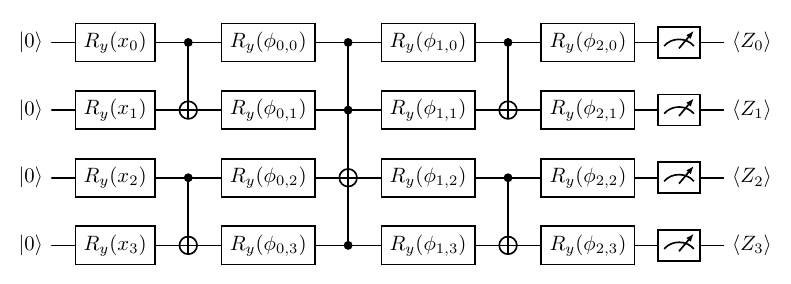}
    \caption{PennyLane ideal quantum circuit}
    \label{fig:pennylane_circuit}
\end{figure}

\subsection{Classical-Quantum PennyLane (Noisy)}
\label{subsec:pennylane_noisy}

To simulate realistic quantum hardware operating in the Noisy Intermediate-Scale Quantum era~\cite{Preskill2018}, we implement noise models based on calibration data from IBM Quantum devices (see Section~\ref{subsubsec:ibm_hardware} for hardware specifications). The PennyLane Noisy architecture employs a density matrix simulator (\texttt{default.mixed}) to model decoherence and gate errors~\cite{Bergholm2018pennylane}.

\paragraph{Noise Implementation}
We compute damping probabilities using exponential decay formulas derived from these coherence times. For single-qubit gates ($t_{\text{gate}}^{(1q)} = 32$ ns) Eq. (\ref{eq:gamma_1q}, \ref{eq:gamma_2q}).

\begin{align}
\gamma_{1q} &= 1 - \exp\left(-\frac{t_{\text{gate}}^{(1q)}}{T_1}\right) \approx 1.28 \times 10^{-4} \label{eq:gamma_1q}\\
\lambda_{1q} &= 1 - \exp\left(-\frac{t_{\text{gate}}^{(1q)}}{T_2}\right) - \gamma_{1q} \approx 8.53 \times 10^{-5} \label{eq:lambda_1q}
\end{align}

Two-qubit CZ gates ($t_{\text{gate}}^{(2q)} = 68$ ns) experience elevated error rates due to longer execution durations \ref{eq:lambda_1q}, \ref{eq:lambda_2q}

\begin{align}
\gamma_{2q} &= 1 - \exp\left(-\frac{t_{\text{gate}}^{(2q)}}{T_1}\right) \approx 2.72 \times 10^{-4} \label{eq:gamma_2q}\\
\lambda_{2q} &= 1 - \exp\left(-\frac{t_{\text{gate}}^{(2q)}}{T_2}\right) - \gamma_{2q} \approx 1.81 \times 10^{-4} \label{eq:lambda_2q}
\end{align}

These probabilities feed into Kraus operator representations for amplitude damping $\mathcal{E}_{\text{AD}}(\gamma)$ and phase damping $\mathcal{E}_{\text{PD}}(\lambda)$. The noisy circuit maintains the identical topology to PennyLane Standard (circular CNOT entanglement, depth $D=3$), but inserts these noise channels after every quantum operation to physically model the relaxation and dephasing inherent to superconducting transmon qubits.

These probabilities feed into Kraus operator representations for amplitude damping $\mathcal{E}_{\text{AD}}(\gamma)$ and phase damping $\mathcal{E}_{\text{PD}}(\lambda)$. The noisy circuit maintains the identical topology to PennyLane Standard (circular CNOT entanglement, depth $D=3$), but inserts these noise channels after every quantum operation to physically model the relaxation and dephasing inherent to superconducting transmon qubits.

\begin{algorithm}
\SetAlgoLined
\KwIn{Features $f$, Noise Params $\Gamma_{1q}, \Lambda_{1q}, \Gamma_{2q}, \Lambda_{2q}$}
Initialize Device `default.mixed`\;
Define Noisy Circuit $U_{noisy}(\theta, \phi)$:\;
\Indp
    $\text{AngleEmbedding}(\theta)$\;
    Apply $\mathcal{E}_{AD}(\Gamma_{1q}) \circ \mathcal{E}_{PD}(\Lambda_{1q})$ on all qubits\;
    \For{$l \gets 1$ \KwTo $D$}{
        \ForEach{CNOT $(q_i, q_j)$ in Entangler}{
            Apply CNOT $(q_i, q_j)$\;
            Apply $\mathcal{E}_{AD}(\Gamma_{2q}) \circ \mathcal{E}_{PD}(\Lambda_{2q})$ on $q_i, q_j$\;
        }
        \ForEach{Rotation $R_Y$ on $q_i$}{
            Apply $R_Y(\phi_{l,i})$\;
            Apply $\mathcal{E}_{AD}(\Gamma_{1q}) \circ \mathcal{E}_{PD}(\Lambda_{1q})$ on $q_i$\;
        }
    }
    \Return $[\langle Z_0 \rangle, \dots, \langle Z_{n-1} \rangle]$\;
\Indm
\caption{Classical-Quantum PennyLane (Noisy).}
\label{alg:pl_noisy}
\end{algorithm}

\subsection{Classical-Quantum Qiskit (Ideal)}
\label{subsec:qiskit_standard}
The Qiskit-based architecture is implemented using the Qiskit Machine Learning framework~\cite{QiskitML}. We distinguish between the ideal and noisy configurations by selecting the appropriate Qiskit Primitives for the NeuralNetwork backend. In the ideal setting, we employ Qiskit's \texttt{EstimatorQNN} to compute exact expectation values of the observable. This choice eliminates stochastic sampling noise, yielding a clean gradient signal for evaluating the theoretical upper bound of the model's expressive capacity. Conversely, for noisy emulation, we use \texttt{SamplerQNN}. This shift is methodologically grounded in the requirement to simulate realistic hardware behavior: while the \texttt{Estimator} provides analytical means, the \texttt{Sampler} reconstructs the quasi-probability distribution from finite counts of bitstrings (shots). This allows the model to incorporate readout errors and shot-noise fluctuations, which are intrinsic to the classification tasks in the NISQ era~\cite{Preskill2018}. By using \texttt{SamplerQNN} in the noisy regime, we ensure that the classification head operates on the same measurement statistics that would be obtained from a physical IBM Quantum processor.

The circuit architecture employs a brick-wall entangling pattern combined with Hadamard initialization. Hadamard gates first create equal superposition states across all qubits using Eq. (\ref{eq:qiskitH}).

\begin{equation}
    |\psi_0\rangle = \bigotimes_{i=0}^{n-1} H^{(i)} |0\rangle = \frac{1}{\sqrt{2^n}} \sum_{j=0}^{2^n-1} |j\rangle
\label{eq:qiskitH}
\end{equation}

where each Hadamard gate $H^{(i)}$ prepares qubit $i$ in uniform superposition, creating an equal probability distribution across all $2^n$ computational basis states. Classical data embedding follows through $R_Y$ rotations~\cite{Schuld2015} using Eq. (\ref{eq:qiskitembedding}).

\begin{equation}
    \hat{U}_{\text{enc}}(\boldsymbol{\theta}) = \prod_{i=0}^{n-1} R_Y^{(i)}(\theta_i)
    \label{eq:qiskitembedding}
\end{equation}

where each rotation $R_Y^{(i)}(\theta_i) = \exp(-i\theta_i Y/2)$ encodes a classical feature $\theta_i$ through rotation around the $Y$ axis.

Variational layers employ parameterized brick-wall entangling blocks~\cite{Mitarai2018}. For each layer $l \in \{0, 1, 2\}$, the transformation combines alternating CNOT patterns with trainable rotations using Eq. (\ref{eq:qiskitvariational}).

\begin{equation}
    \hat{U}_{\text{var}}^{(l)} = \hat{U}_{\text{rot}}(\boldsymbol{\phi}_l) \cdot \hat{U}_{\text{ent}}^{\text{odd}} \cdot \hat{U}_{\text{ent}}^{\text{even}}
    \label{eq:qiskitvariational}
\end{equation}

The even CNOT layer entangles adjacent qubit pairs $(0,1)$ and $(2,3)$ using Eq. (\ref{eq:qiskiteven}).

\begin{equation}
    \hat{U}_{\text{ent}}^{\text{even}} = \text{CNOT}_{0,1} \cdot \text{CNOT}_{2,3}
    \label{eq:qiskiteven}
\end{equation}

while the odd layer targets the intermediate pair $(1,2)$ using Eq. (\ref{eq:qiskitodd})

\begin{equation}
    \hat{U}_{\text{ent}}^{\text{odd}} = \text{CNOT}_{1,2}
    \label{eq:qiskitodd}
\end{equation}

This alternating pattern ensures efficient entanglement distribution~\cite{Sim2019}, thereby establishing connectivity between all qubit pairs within the layered structure. Trainable rotations following entanglement apply parameterized $R_Y$ gates using Eq. (\ref{eq:qiskitentanglement}).

\begin{equation}
    \hat{U}_{\text{rot}}(\boldsymbol{\phi}_l) = \prod_{i=0}^{n-1} R_Y^{(i)}(\phi_{l,i})
    \label{eq:qiskitentanglement}
\end{equation}

The complete circuit combines initialization, encoding, and variational layers as specified in Eq. (\ref{eq:qiskitencoding}).

\begin{equation}
    \hat{U}_{\text{Qiskit}}(\boldsymbol{\theta}, \{\boldsymbol{\phi}_l\}) = \prod_{l=0}^{D-1} \hat{U}_{\text{var}}^{(l)}(\boldsymbol{\phi}_l) \cdot \hat{U}_{\text{enc}}(\boldsymbol{\theta}) \cdot \prod_{i=0}^{n-1} H^{(i)}
    \label{eq:qiskitencoding}
\end{equation}

with total trainable parameters reaching $n \times D = 4 \times 3 = 12$.

\begin{figure}[htbp]
    \centering
    \includegraphics[width=0.7\textwidth]{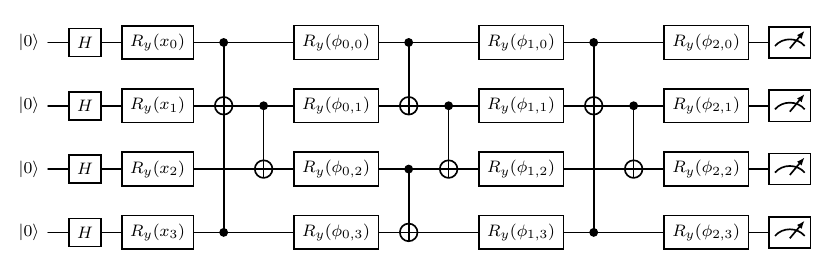}
    \caption{Qiskit quantum circuit employing brick-wall entangling pattern (4 qubits, depth=3)}
    \label{fig:qiskit_circuit}
\end{figure}

All $n$ qubits undergo measurement in the computational basis, producing binary strings $\mathbf{b} \in \{0,1\}^n$. The interpretation function maps binary strings to class labels by computing the Hamming weight using Eq. (\ref{eq:qiskitmap}).
 
\begin{equation}
    f_{\text{interpret}}(\mathbf{b}) = \text{HammingWeight}(\mathbf{b}) \mod C
    \label{eq:qiskitmap}
\end{equation}

where $\text{HammingWeight}(\mathbf{b}) = \sum_{i=0}^{n-1} b_i$ counts the number of ones in the binary string and $C$ denotes the number of classes. The \texttt{SamplerQNN} outputs probability distributions over classes by aggregating binary strings that map to identical labels using Eq. (\ref{eq:qiskitmaplabel}).

\begin{equation}
    P(c) = \sum_{\mathbf{b} : f(\mathbf{b})=c} |\langle \mathbf{b} | \psi(\boldsymbol{\theta}, \boldsymbol{\phi}) \rangle|^2
    \label{eq:qiskitmaplabel}
\end{equation}

The standard noiseless version employs Qiskit's \texttt{Sampler} primitive with exact statevector simulation~\cite{Qiskit}, yielding deterministic outputs without shot noise or decoherence.

\begin{algorithm}
\SetAlgoLined
\KwIn{Features $f$, Qubits $n=4$, Depth $D=3$}
Initialize `EstimatorQNN` with Gradient support\;
Define Circuit $QC$:\;
\Indp
    Apply Hadamard to all $q$\;
    Apply $R_Y(\theta)$ \tcp*{Feature Map}
    \For{$l \gets 1$ \KwTo $D$}{
        Apply Brick-Wall CNOTs\;
        Apply $R_Y(\phi_l)$ \tcp*{Variational}
    }
\Indm
Define Observables $O = \{ZIII, IZII, \dots\}$\;
\textbf{Forward Pass:}\;
\Indp
    $\theta \leftarrow \tanh(W_{in} \cdot f) \cdot \frac{\pi}{2}$\;
    $exp\_vals \leftarrow \text{Estimator.run}(QC, \theta, O)$\;
    \Return $exp\_vals$\;
\Indm
\caption{Classical-Quantum Qiskit (Ideal)}
\label{alg:qiskit_standard}
\end{algorithm}

\subsection{Classical-Quantum Qiskit (Noisy)}
\label{subsec:qiskit_noisy}
The Qiskit Noisy architecture employs the Qiskit Aer simulator~\cite{QiskitAer}, incorporating depolarizing noise channels and finite-shot measurements. 

\paragraph{Justification for noise model selection}
To ensure a fair comparison between frameworks, both PennyLane and Qiskit implementations use identical underlying noise parameters derived from Table~\ref{tab:ibm_calibration_unified}. While PennyLane implements \emph{amplitude and phase damping} channels (physically accurate for superconducting transmon qubits), Qiskit employs \emph{thermal relaxation combined with depolarizing noise}. Both approaches are calibrated to achieve equivalent circuit fidelity for depth-3, 4-qubit circuits, as measured by Eq. (\ref{eq:qiskitfidelity}).
\begin{equation}
F_{\text{target}} \approx 0.94\text{--}0.96
\label{eq:qiskitfidelity}
\end{equation}
This high fidelity range reflects the significantly improved performance of Heron r2 processors.

\paragraph{Qiskit Noise Calibration}
Qiskit implements thermal relaxation errors using the same $T_1$ and $T_2$ values from Table~\ref{tab:ibm_calibration_unified}, combined with depolarizing channels calibrated to match Heron r2 gate error rates:

\begin{align}
p_{1q} &= 0.0002 \quad (0.02\% \text{ single-qubit error rate}) \\
p_{2q} &= 0.005 \quad (0.5\% \text{ two-qubit error rate})
\end{align}

These values are identical to the gate error rates used in PennyLane, ensuring equivalent noise intensity across frameworks.

\paragraph{Fidelity equivalence verification}
With unified noise parameters based on the Heron r2 specifications, both frameworks aim for equivalent-circuit fidelity. For a depth-3, 4-qubit circuit (approximately 28 total gates), we estimate using Eq. (\ref{eq:qiskitdepolarizing}):
\begin{equation}
F_{\text{PL}} \approx F_{\text{QK}} \approx 0.94\text{--}0.96
\label{eq:qiskitdepolarizing}
\end{equation}
This higher fidelity range reflects the improved coherence and gate quality of Heron processors compared to earlier generations. The equivalence ensures that observed performance differences reflect framework-specific factors rather than noise intensity discrepancies.

\paragraph{Shot Noise and Sampling}
Unlike the density matrix simulation used in PennyLane, Qiskit Noisy employs finite-shot sampling ($N_{\text{shots}} = 1024$). This introduces statistical noise independent of gate errors, as given by Eq. (\ref{eq:qiskitnoise}).

\begin{equation}
\sigma_{\text{shot}}[\hat{P}(b)] \approx \frac{1}{\sqrt{N_{\text{shots}}}} \approx 3.1\%
\label{eq:qiskitnoise}
\end{equation}

This sampling variance adds to the depolarizing noise, creating a two-fold stochasticity (gate errors + measurement uncertainty). We acknowledge this asymmetry as inherent to the framework architectures: PennyLane computes expectation values analytically via trace operations, whereas Qiskit samples computational-basis states, mimicking hardware behavior. This distinction allows us to compare an upper-bound performance limit (PennyLane) against a realistic deployment scenario (Qiskit).

\begin{algorithm}
\SetAlgoLined
\KwIn{Features $f$, Backend $B$, Shots $N=1024$}
\KwOut{Class Probabilities}
Build `NoiseModel` from backend $B$ calibration\;
Initialize `SamplerQNN` with $N$ shots and NoiseModel\;
Define Interpret Function $I(b) = \text{Hamming}(b) \pmod C$\;
\textbf{Forward Pass:}\;
\Indp
    $\theta \leftarrow \text{ScaleFeatures}(f)$\;
    Sample binary strings $b \sim |\psi(\theta, \phi)|^2$ ($N$ times)\;
    Compute raw counts $\{b: \text{count}\}$\;
    Map counts to classes using $I(b)$\;
    $P(c) \leftarrow \frac{\sum_{b: I(b)=c} \text{count}(b)}{N}$\;
    \Return $P(c)$\;
\Indm
\caption{Classical-Quantum Qiskit (Noisy).}
\label{alg:qiskit_noisy}
\end{algorithm}

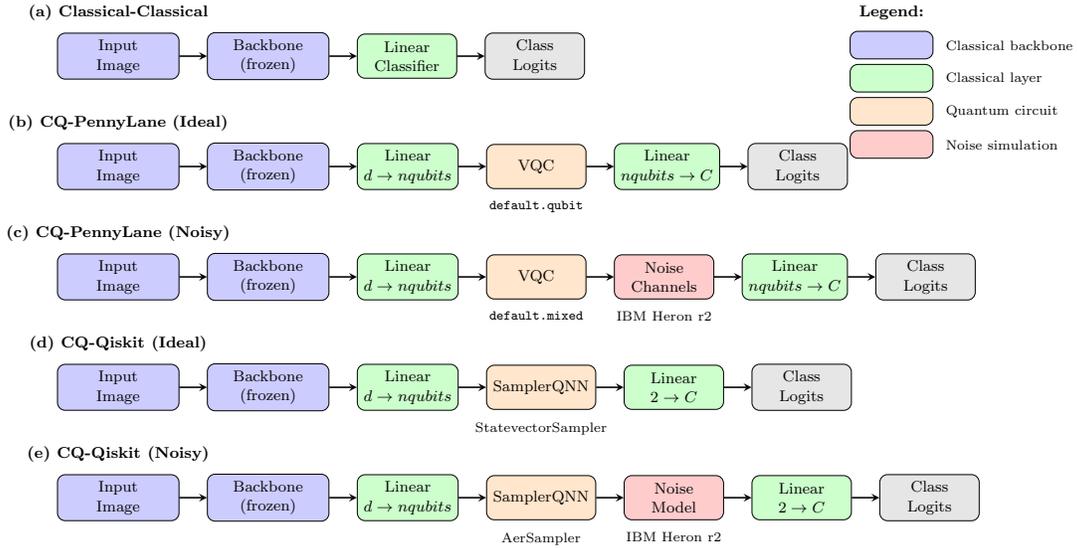
\begin{figure}
\centering
\resizebox{\textwidth}{!}{%
\begin{tikzpicture}[
    node distance=0.3cm and 0.5cm,
    box/.style={rectangle, draw, rounded corners, minimum width=1.8cm, minimum height=0.8cm, align=center, font=\footnotesize},
    backbone/.style={box, fill=blue!20, minimum width=2.2cm},
    classical/.style={box, fill=green!20},
    quantum/.style={box, fill=orange!20},
    noise/.style={box, fill=red!20},
    output/.style={box, fill=gray!20},
    arrow/.style={->, thick, >=stealth},
    title/.style={font=\footnotesize\bfseries, align=center},
    label/.style={font=\scriptsize, align=center},
]

\def\rowsep{2.0}
\def\colsep{2.5}

\node[title] at (0, 0.8) {(a) Classical-Classical};
\node[backbone] (img1) at (0, 0) {Input\\Image};
\node[backbone, right=of img1] (cnn1) {Backbone\\(frozen)};
\node[classical, right=of cnn1] (fc1) {Linear\\Classifier};
\node[output, right=of fc1] (out1) {Class\\Logits};
\draw[arrow] (img1) -- (cnn1);
\draw[arrow] (cnn1) -- (fc1);
\draw[arrow] (fc1) -- (out1);

\node[title] at (0, -\rowsep+0.8) {(b) CQ-PennyLane (Ideal)};
\node[backbone] (img2) at (0, -\rowsep) {Input\\Image};
\node[backbone, right=of img2] (cnn2) {Backbone\\(frozen)};
\node[classical, right=of cnn2] (pre2) {Linear\\$d \to nqubits$};
\node[quantum, right=of pre2] (qc2) {VQC};
\node[classical, right=of qc2] (post2) {Linear\\$nqubits \to C$};
\node[output, right=of post2] (out2) {Class\\Logits};
\draw[arrow] (img2) -- (cnn2);
\draw[arrow] (cnn2) -- (pre2);
\draw[arrow] (pre2) -- (qc2);
\draw[arrow] (qc2) -- (post2);
\draw[arrow] (post2) -- (out2);
\node[label, below=0.1cm of qc2] {\texttt{default.qubit}};

\node[title] at (0, -2*\rowsep+0.8) {(c) CQ-PennyLane (Noisy)};
\node[backbone] (img3) at (0, -2*\rowsep) {Input\\Image};
\node[backbone, right=of img3] (cnn3) {Backbone\\(frozen)};
\node[classical, right=of cnn3] (pre3) {Linear\\$d \to nqubits$};
\node[quantum, right=of pre3] (qc3) {VQC};
\node[noise, right=of qc3] (noise3) {Noise\\Channels};
\node[classical, right=of noise3] (post3) {Linear\\$nqubits \to C$};
\node[output, right=of post3] (out3) {Class\\Logits};
\draw[arrow] (img3) -- (cnn3);
\draw[arrow] (cnn3) -- (pre3);
\draw[arrow] (pre3) -- (qc3);
\draw[arrow] (qc3) -- (noise3);
\draw[arrow] (noise3) -- (post3);
\draw[arrow] (post3) -- (out3);
\node[label, below=0.1cm of qc3] {\texttt{default.mixed}};
\node[label, below=0.1cm of noise3] {IBM Heron r2};

\node[title] at (0, -3*\rowsep+0.8) {(d) CQ-Qiskit (Ideal)};
\node[backbone] (img4) at (0, -3*\rowsep) {Input\\Image};
\node[backbone, right=of img4] (cnn4) {Backbone\\(frozen)};
\node[classical, right=of cnn4] (pre4) {Linear\\$d \to nqubits$};
\node[quantum, right=of pre4] (qc4) {SamplerQNN};
\node[classical, right=of qc4] (post4) {Linear\\$2 \to C$};
\node[output, right=of post4] (out4) {Class\\Logits};
\draw[arrow] (img4) -- (cnn4);
\draw[arrow] (cnn4) -- (pre4);
\draw[arrow] (pre4) -- (qc4);
\draw[arrow] (qc4) -- (post4);
\draw[arrow] (post4) -- (out4);
\node[label, below=0.1cm of qc4] {StatevectorSampler};

\node[title] at (0, -4*\rowsep+0.8) {(e) CQ-Qiskit (Noisy)};
\node[backbone] (img5) at (0, -4*\rowsep) {Input\\Image};
\node[backbone, right=of img5] (cnn5) {Backbone\\(frozen)};
\node[classical, right=of cnn5] (pre5) {Linear\\$d \to nqubits$};
\node[quantum, right=of pre5] (qc5) {SamplerQNN};
\node[noise, right=of qc5] (noise5) {Noise\\Model};
\node[classical, right=of noise5] (post5) {Linear\\$2 \to C$};
\node[output, right=of post5] (out5) {Class\\Logits};
\draw[arrow] (img5) -- (cnn5);
\draw[arrow] (cnn5) -- (pre5);
\draw[arrow] (pre5) -- (qc5);
\draw[arrow] (qc5) -- (noise5);
\draw[arrow] (noise5) -- (post5);
\draw[arrow] (post5) -- (out5);
\node[label, below=0.1cm of qc5] {AerSampler};
\node[label, below=0.1cm of noise5] {IBM Heron r2};

\node[title] at (14, 0.8) {Legend:};
\node[backbone, minimum width=1.5cm, minimum height=0.5cm] (leg1) at (14, 0.2) {};
\node[right=0.1cm of leg1, font=\scriptsize] {Classical backbone};
\node[classical, minimum width=1.5cm, minimum height=0.5cm] (leg2) at (14, -0.4) {};
\node[right=0.1cm of leg2, font=\scriptsize] {Classical layer};
\node[quantum, minimum width=1.5cm, minimum height=0.5cm] (leg3) at (14, -1.0) {};
\node[right=0.1cm of leg3, font=\scriptsize] {Quantum circuit};
\node[noise, minimum width=1.5cm, minimum height=0.5cm] (leg4) at (14, -1.6) {};
\node[right=0.1cm of leg4, font=\scriptsize] {Noise simulation};

\end{tikzpicture}
}
\caption{Overview of the five evaluated architectures: classical baseline and four hybrid classical-quantum variants.}
\label{fig:architecture_overview}
\end{figure}

As a summary of the methodological setup, Figure \ref{fig:architecture_overview} presents a schematic overview of the five evaluated architectures, highlighting the shared feature extraction stage and the classical and quantum transfer learning heads considered in this work.



\section{Results}
\label{sec:results}
In this section, we describe the datasets selected for the study (Section \ref{sec:Data}). We also detail the metric used to evaluate the models (Section \ref{sec:Metrics}). The complete experimental setup is also presented, including the pretrained models employed, the hardware and software environment, and all relevant hyperparameters. Additionally, we outline the configuration of the quantum circuit (Section \ref{sec:Experimental}). Finally, we compare the obtained results in terms of both execution time and accuracy (Section \ref{subsec:result_discussion}).

\subsection{Datasets}\label{sec:Data}
We evaluate four binary image datasets of heterogeneous difficulty: Hymenoptera (ants vs. bees), Brain Tumor MRI (tumor vs. no tumor), Cats vs. Dogs, and Solar Dust (defect vs. normal). All images are resized to $224\times224$ and preprocessed with ImageNet statistics. The training set is split 80\% for training and 20\% for validation, the held-out test set is used only for final reporting.

Table~\ref{tab:dataset_summary} provides a comparative overview of the four datasets, highlighting the diversity in scale, domain, and complexity.

\begin{table}[htbp]
\centering
\caption{Comprehensive dataset characteristics summary.}
\label{tab:dataset_summary}
\begin{tabular}{lccccl}
\hline
\textbf{Dataset} & \textbf{Train} & \textbf{Test} & \textbf{Total} & \textbf{Difficulty} & \textbf{Domain} \\
\hline
Hymenoptera~\cite{TheDataSith2021} & 245 & 153 & 398 & Medium & Entomology \\
Brain Tumor~\cite{Nickparvar2021} & 2000 & 400 & 2400 & High & Medical Imaging \\
Cats vs. Dogs~\cite{Kaggle2013} & 3000 & 600 & 3600 & Medium-Low & General Vision \\
Solar Dust~\cite{Garladinne2023} & 240 & 60 & 300 & High & Industrial \\
\hline
\end{tabular}
\end{table}

\subsection{Metrics}\label{sec:Metrics}
Our metric is test accuracy, defined as Eq. (\ref{eq:acc}).
\begin{equation}
  	\text{Acc} = \frac{1}{N} \sum_{i=1}^{N} 1(\hat{y}_i = y_i),
    \label{eq:acc}
\end{equation}
with $N$ test samples, true labels $y_i$ and predictions $\hat{y}_i$, where 
$1(\cdot)$ denotes the indicator function. We track training time (seconds) for efficiency $t_s$.

Unless stated otherwise, all runs reported here use 10 epochs and are executed on the CPU. Accuracy values are reported in $[0,1]$. For readability, we discuss percentage points when comparing approaches.

\subsection{Experimental Setup}\label{sec:Experimental}
Across all experiments, we freeze the convolutional backbone and train only the head (classical linear layer or quantum head). Common hyperparameters: Adam ($\eta=10^{-3}$, batch size 16), 10 epochs. Quantum settings: 4 qubits and depth 3. For Qiskit Standard (SamplerQNN), we use 1024 shots and a depolarizing noise model ($p_{1q}=0.0002$, $p_{2q}=0.005$), calibrated to match Heron r2 gate error rates; the no-noise variant disables both shots and noise. PennyLane Standard uses \texttt{default.qubit} (noiseless). PennyLane Noisy follows realistic device-like damping/dephasing as in Section~\ref{subsec:pennylane_noisy}.

\subsubsection{CNNs Pretrained Backbones}
We evaluate four backbones: ResNet-18~\cite{He2016} contains 11.7M parameters, with residual connections, producing 512-dimensional features. MobileNetV2~\cite{Sandler2018mobilenetv2} employs 3.5M parameters using depthwise separable convolutions, yielding 1280-dimensional features. EfficientNet-B0~\cite{Tan2019} integrates 5.3M parameters through compound scaling, generating 1280-dimensional features. RegNet-X-400MF~\cite{Radosavovic2020designing} comprises 5.2M parameters through a NAS-designed architecture, extracting 400-dimensional features.

\subsubsection{Hardware and Software Environment}

All experiments were executed on a standardized computing platform, summarized in Table~\ref{tab:hardware_software}.

\begin{table}[htbp]
\centering
\caption{Hardware and software environment used in all experiments.}
\label{tab:hardware_software}
\begin{tabular}{ll}
\toprule
\textbf{Component} & \textbf{Specification} \\
\midrule
CPU & AMD Ryzen 5 3600 (3.6 GHz, 6 cores) \\
RAM & 24 GB DDR4-3200 \\
Operating system & Windows 10 Pro \\
Python & 3.13.5 \\
PyTorch & 2.1.0+cpu \\
TorchVision & 0.16.0 \\
PennyLane & 0.35 \\
Qiskit & $\geq$1.0.0, $<$2.0.0 \\
Qiskit Machine Learning & 0.8.0 \\
Qiskit Aer & 0.13.1 \\
\bottomrule
\end{tabular}
\end{table}

CPU-only execution was deliberately chosen to eliminate GPU-related variability and ensure consistent timing measurements across quantum and classical components. All simulations were run single-threaded for quantum circuit operations, whereas the classical backbones relied on PyTorch's default CPU parallelization.

\subsubsection{Training Hyperparameters}

We adopt identical hyperparameters across all five architectures to isolate the impact of quantum components:

\begin{table}[htbp]
\centering
\caption{Unified training hyperparameters for all architectures.}
\label{tab:hyperparams}

\begin{tabular}{ll}
\toprule
\textbf{Parameter} & \textbf{Value} \\
\midrule
Optimizer & Adam \\
Learning rate ($\eta$) & $1 \times 10^{-3}$ \\
Adam $\beta_1$ & 0.9 \\
Adam $\beta_2$ & 0.999 \\
Adam $\epsilon$ & $1 \times 10^{-8}$ \\
Weight decay & 0 (frozen backbone) \\
LR scheduler & StepLR \\
Step size & 3 epochs \\
Gamma ($\gamma$) & 0.9 \\
Batch size & 16 \\
Total epochs & 10 \\
Gradient clipping & None (stable gradients) \\
\midrule
\multicolumn{2}{l}{\textit{Quantum-specific settings:}} \\
\midrule
Number of qubits ($n$) & 4 \\
Circuit depth ($D$) & 3 \\
PennyLane device (clean) & \texttt{default.qubit} \\
PennyLane device (noisy) & \texttt{default.mixed} \\
PennyLane gradient & Parameter-shift rule \\
Qiskit shots & 1024 \\
Qiskit gradient & SPSA (2 evaluations) \\
\bottomrule
\end{tabular}
\end{table}

The StepLR scheduler reduces the learning rate by a factor $\gamma=0.9$ every 3 epochs, providing gentle annealing without aggressive decay. This conservative schedule proved effective across all datasets without requiring dataset-specific tuning.

\subsubsection{Quantum Circuit Configuration}
Both PennyLane and Qiskit variants employ shallow variational quantum circuits with four qubits and depth three, yielding a total of 12 trainable quantum parameters ($n \times D = 4 \times 3 = 12$). This configuration was deliberately chosen to balance expressive capacity with noise robustness, reflecting practical constraints imposed by current NISQ hardware. Deeper or wider circuits were avoided to limit decoherence and ensure stable optimization during training.

Gradient computation follows framework-specific best practices. In the PennyLane implementations, gradients are evaluated analytically using the parameter-shift rule~\cite{Mitarai2018}, which computes exact derivatives through a fixed number of circuit evaluations per parameter. This approach makes deterministic gradients and avoids additional stochasticity beyond that induced by quantum noise in the noisy simulations.

By contrast, Qiskit-based models rely on Simultaneous Perturbation Stochastic Approximation (SPSA) for gradient estimation \cite{Spall1992, Guerreschi2020}. SPSA approximates gradients via finite differences along randomly sampled perturbation directions, requiring only two circuit evaluations per optimization step. While this introduces controlled stochastic noise into the optimization process, it is well-suited to shot-based quantum execution. It aligns with standard practice in variational quantum algorithms implemented in the Qiskit ecosystem.

Although these gradient computation strategies differ substantially, they are both consistent with the design philosophies of their respective frameworks \cite{Bergholm2018pennylane, Qiskit}. Pennylane is architected around hardware-compatible automatic differentiation, where the Parameter-Shift rule provides unbiased gradients at the cost of 2n circuit evaluations \cite{Bergholm2018pennylane}. Conversely, Qiskit's common integration with SPSA addresses the high-variance noise floor of current IBM hardware by using a gradient-free stochastic approximation that remains effective under strict shot counts \cite{Guerreschi2020,amaro2022filtering}.

Consequently, the observed differences in optimization behavior should be interpreted as intrinsic to the underlying toolchains rather than as artifacts of ad hoc methodological choices.

\subsubsection{IBM Quantum Hardware Configuration}
\label{subsubsec:ibm_hardware}

To validate our simulation results, we conducted experiments on real IBM Quantum hardware accessed through IBM Quantum Cloud services. Both noisy simulations and real hardware experiments use noise parameters calibrated from IBM Heron r2 processors, ensuring consistency across evaluation settings.

\paragraph{Hardware Specifications}
Noise parameters for simulation are based on publicly available specifications for IBM Quantum Heron r2 processors (2024--2025). These processors feature improved coherence times and gate fidelities compared to earlier Falcon-generation devices. Table~\ref{tab:ibm_calibration_unified} presents the unified calibration parameters used across both PennyLane and Qiskit implementations.

\begin{table}[htbp]
\centering
\caption{IBM Heron r2 processor specifications used for noise calibration and real hardware experiments.}
\label{tab:ibm_calibration_unified}
\begin{tabular}{lcc}
\toprule
\textbf{Parameter} & \textbf{Value} & \textbf{Usage} \\
\midrule
\multicolumn{3}{l}{\textit{Coherence properties:}} \\
$T_1$ (energy relaxation) & 250 $\mu$s & PL damping \\
$T_2$ (dephasing time) & 150 $\mu$s & PL dephasing \\
\midrule
\multicolumn{3}{l}{\textit{Gate timing:}} \\
Single-qubit gate time & 32 ns & PL noise calc. \\
Two-qubit gate time (CZ) & 68 ns & PL noise calc. \\
\midrule
\multicolumn{3}{l}{\textit{Error rates:}} \\
Single-qubit gate error ($p_{1q}$) & 0.02\% & Qiskit depol. \\
Two-qubit gate error ($p_{2q}$) & 0.5\% & Qiskit depol. \\
Readout error (SPAM) & 1.2\% & Both \\
\bottomrule
\end{tabular}
\end{table}

For PennyLane noisy simulations, we compute amplitude and phase damping probabilities from coherence times using exponential decay formulas (Eqs.~\ref{eq:gamma_1q}--\ref{eq:lambda_2q}). For Qiskit noisy simulations, we apply depolarizing channels with error rates matching the gate error specifications. This unified calibration ensures that both frameworks experience the same noise intensity, enabling a fair cross-platform comparison.

\paragraph{Real Hardware Configuration}
For real QPU experiments, we executed on \texttt{ibm\_torino}, a 133-qubit IBM Heron r2 processor. Table~\ref{tab:real_hw_config} summarizes the execution configuration.

\begin{table}[htbp]
\centering
\caption{Real quantum hardware execution configuration.}
\label{tab:real_hw_config}
\begin{tabular}{ll}
\toprule
\textbf{Parameter} & \textbf{Value} \\
\midrule
Backend & \texttt{ibm\_torino} \\
Processor type & IBM Heron r2 \\
Total qubits & 133 \\
Qubits used & 4 \\
Circuit depth (transpiled) & $\sim$49 \\
Shots per circuit & 100 \\
Transpilation level & 1 \\
\bottomrule
\end{tabular}
\end{table}

The transpilation process maps the logical 4-qubit circuit to physical qubits on the device topology, inserting SWAP gates as necessary. The resulting transpiled circuit depth of approximately 49 gates reflects the overhead introduced by hardware connectivity constraints.

\paragraph{Circuit Batching and Gradient Estimation}
Real hardware execution employs Qiskit's \texttt{SamplerV2} primitive for efficient circuit batching, aggregating all samples in a training batch into a single job submission. For gradient computation, we use Simultaneous Perturbation Stochastic Approximation (SPSA)~\cite{Spall1992}, which approximates gradients using only two circuit evaluations per step (Eq.~\ref{eq:SPSA}):
\begin{equation}
    \label{eq:SPSA}
    \hat{g}_k = \frac{L(\boldsymbol{\theta} + c_k \boldsymbol{\Delta}_k) - L(\boldsymbol{\theta} - c_k \boldsymbol{\Delta}_k)}{2c_k} \boldsymbol{\Delta}_k
\end{equation}
where $\boldsymbol{\Delta}_k \in \{-1, +1\}^P$ is a random Bernoulli perturbation vector and $c_k=0.3$ is the perturbation magnitude. This reduces hardware evaluations from 72 (parameter-shift) to 3 per batch (1 forward + 2 SPSA).

\paragraph{Training Protocol Adaptations}
Due to QPU access constraints (queue wait times, daily quotas, execution costs), real hardware experiments use: 5 epochs (vs. 10 in simulation), Hymenoptera dataset only (moderate size balancing significance with time), 100 shots per circuit, and SPSA learning rate $\eta=0.1$. These adaptations reflect practical realities of NISQ hardware and should be considered when interpreting the results in Table~\ref{tab:hym_detail}.

\subsection{Result Discussion}
\label{subsec:result_discussion}

We report the mean training time per run (in seconds), averaged over the four backbones for 10 epochs, as summarized in Table~\ref{tab:time_avg}.
\subsubsection{Hymenoptera}
\label{subsubsec:hymenoptera}

Table~\ref{tab:hym_detail} presents the per-backbone results on the Hymenoptera dataset demonstrate the most advantage over the classical baseline.

\begin{table}[htbp]
\centering
\caption{Hymenoptera: Test accuracy and (training time in seconds) per backbone.}
\label{tab:hym_detail}
\resizebox{\textwidth}{!}{%
\begin{tabular}{lccccccc}
\toprule
\textbf{Backbone} & \textbf{CC} & \textbf{PennyLane} & \textbf{PL-Noisy} & \textbf{PL-Real} & \textbf{Qiskit} & \textbf{Qiskit-Noisy} & \textbf{Qiskit-Real} \\
\midrule
ResNet18        & 0.8589 (148.9) & \textbf{0.9744} (107.2) & 0.9615 (178.4) & 0.8421 (704) & 0.9359 (305.7) & 0.9295 (352.1) & 0.9487 (559) \\
MobileNetV2     & 0.8846 (136.7) & 0.9487 (80.7)  & 0.9423 (142.8) & 0.8684 (685) & \textbf{0.9744} (288.2) & 0.9551 (331.6) & 0.9103 (681)\\
EfficientNet-B0 & 0.9231 (178.9) & \textbf{0.9487} (107.9) & 0.9359 (168.5) & 0.6579 (682) & 0.9231 (297.4) & 0.9103 (348.9) & 0.7179 (545)\\
RegNet-X-400MF  & 0.8590 (81.0)  & \textbf{0.9231} (83.4)  & 0.9103 (128.7) & 0.7895 (673) & 0.9103 (270.2) & 0.8974 (312.4) & 0.8590 (511)\\
\bottomrule
\end{tabular}
}
\end{table}

This dataset exhibits the largest accuracy gap between quantum-hybrid and classical approaches. PennyLane-Standard with ResNet18 achieves 0.9744, outperforming the best classical result (EfficientNet-B0, 0.9231) by over 5\%. The same-backbone comparison is even more striking: PennyLane improves ResNet18 from 0.8589 to 0.9744, MobileNetV2 from 0.8846 to 0.9487, and RegNet-X-400MF from 0.8590 to 0.9231. Qiskit-Standard matches PennyLane's peak performance with MobileNetV2 (0.9744) and also surpasses the classical baseline across all backbones. Under noise, PL-Noisy degrades by 0.6--1.3\% relative to PennyLane ideal, while Qiskit-Noisy shows a wider spread of 0.6--1.9\% relative to Qiskit ideal; nevertheless, every noisy variant still exceeds the best classical result. The consistent quantum advantage across all four backbones suggests that, on small datasets with moderate intra-class variability, the compact 12-parameter quantum head captures discriminative features that the frozen classical layer alone fails to capture.

Averaged over the four backbones, PennyLane-Standard is the fastest approach (94.8\,s), reducing training time by 30\% with respect to the classical baseline (136.4\,s). PL-Noisy rises to 154.6\,s (a 63\% overhead over PL ideal), placing it 13\% above the classical cost. Qiskit-Standard (290.4\,s) and Qiskit-Noisy (336.3\,s) are approximately $2.1\times$ and $2.5\times$ slower than the classical baseline, respectively; the noise model adds 16\% on top of ideal Qiskit. The overall pattern mirrors Cats vs.\\ Dogs in relative terms, though absolute times are roughly $7\times$ shorter owing to the much smaller dataset (398 vs.\\ 3\,600 images).

Hymenoptera is the only dataset evaluated on real IBM Quantum processors (\texttt{ibm\_torino}, 5 epochs, 100 shots), enabling a direct comparison across the full simulation-to-hardware pipeline. Qiskit-Real with MobileNetV2 achieves 94.87\%, remarkably close to its ideal simulation counterpart (97.44\%) and only 0.6\% below the noisy simulation (95.51\%), validating the predictive value of the calibrated noise model. ResNet18 follows a similar trend (Qiskit-Real 94.87\% vs.\ Qiskit-Noisy 92.95\%), actually exceeding the noisy simulation---likely due to favourable hardware conditions during the run. However, EfficientNet-B0 suffers a sharp drop on real hardware (71.79\% vs.\ 91.03\% noisy), suggesting that its higher-dimensional feature space (1\,280-d) is more sensitive to the compounding effects of transpilation overhead and SPSA gradient noise on physical qubits. PennyLane-Real shows a systematically larger accuracy gap relative to its ideal simulation: MobileNetV2 drops from 94.87\% to 86.84\% ($-$8.0\%) and EfficientNet-B0 from 94.87\% to 65.79\% ($-$29.1\%). This wider degradation is attributed to the transition from the analytic parameter-shift gradient used in simulation to SPSA on hardware, which introduces additional stochasticity that the PennyLane-trained weights were not exposed to during ideal optimization. Regarding training time, real hardware runs average 670--690\,s for PL-Real and 511--681\,s for Qiskit-Real (5 epochs), approximately $5\times$ slower per epoch than the corresponding noisy simulations, primarily due to job submission overhead and queue latency rather than circuit execution itself. Despite these constraints, the Qiskit-Real results confirm that quantum transfer learning is practically deployable on current NISQ devices, with MobileNetV2 and ResNet18 achieving accuracy levels competitive with noisy simulation at a manageable time cost.

\begin{table}[htbp]
\centering
\caption{Average training time (s) per dataset and architecture (10 epochs, CPU).}
\label{tab:time_avg}
\begin{tabular}{lccccc}
\toprule
\textbf{Dataset} & \textbf{CC} & \textbf{PennyLane} & \textbf{PL-Noisy} & \textbf{Qiskit} & \textbf{Qiskit-Noisy}\\
\midrule
Hymenoptera    & 136.4 & \textbf{94.8}  & 154.6  & 290.4 & 336.3 \\
Brain Tumor    & 665.7 & \textbf{442.2} & 728.3 & 1458.4 & 1685.4 \\
Cats vs. Dogs  & 1024.5 & \textbf{665.2} & 1102.0 & 2169.1 & 2507.2 \\
Solar Dust     & 309.5 & \textbf{240.1} & 397.2 & 544.6 & 629.6 \\
\bottomrule
\end{tabular}
\end{table}

\subsubsection{Brain Tumor}
\label{subsubsec:brain_tumor}

Table~\ref{tab:brain_detail} presents the per-backbone results on Brain Tumor.

\begin{table}[htbp]
\centering
\caption{Brain Tumor: Test accuracy and training time (in seconds) per backbone.}
\label{tab:brain_detail}
\resizebox{\textwidth}{!}{%
\begin{tabular}{lccccc}
\toprule
\textbf{Backbone} & \textbf{CC} & \textbf{PennyLane} & \textbf{PL-Noisy} & \textbf{Qiskit} & \textbf{Qiskit-Noisy} \\
\midrule
ResNet18        & 0.9725 (736.8) & 0.9500 (541.9) & 0.9425 (892.7) & \textbf{0.9850} (1510.7) & 0.9775 (1745.3) \\
MobileNetV2     & \textbf{0.9825} (639.3) & 0.9125 (418.1) & 0.9050 (687.4) & 0.9725 (1447.1) & 0.9600 (1672.8) \\
EfficientNet-B0 & \textbf{0.9725} (891.9) & 0.9625 (476.0) & 0.9525 (784.2) & 0.9600 (1515.8) & 0.9500 (1751.9) \\
RegNet-X-400MF  & \textbf{0.9900} (394.9) & 0.9400 (332.9) & 0.9350 (548.7) & 0.9450 (1359.9) & 0.9375 (1571.4) \\
\bottomrule
\end{tabular}
}
\end{table}

Unlike the previous dataset, the classical baseline dominates here. CC-RegNet-X-400MF achieves the highest overall accuracy (0.9900), and the classical head also outperforms MobileNetV2 (0.9825) and EfficientNet-B0 (0.9725). The only quantum result that surpasses a classical counterpart is Qiskit-Standard with ResNet18 (0.9850 vs.\\ 0.9725). PennyLane-Standard underperforms the classical baseline on all backbones, with the gap ranging from 1.0\% on EfficientNet-B0 to 7.0\% on MobileNetV2, suggesting that the statevector-based quantum head does not add value when the classical features are already highly separable. Qiskit-Standard remains closer to the classical level, likely because its shot-based sampling provides implicit regularization. Noise degrades PL-Noisy by 0.5--1.0\% and Qiskit-Noisy by 0.75--1.25\% relative to their ideal counterparts, the smallest drops observed across all datasets, consistent with the hypothesis that high baseline accuracy leaves limited room for noise-induced errors to change the classification outcome.

PennyLane-Standard remains the fastest variant (442.2\,s on average), reducing training time by 34\% compared with the classical baseline (665.7\,s). PL-Noisy reaches 728.3\,s (a 65\% overhead over PL ideal), approximately 9\% above the classical cost. Qiskit-Standard (1\,458.4\,s) and Qiskit-Noisy (1\,685.4\,s) are $2.2\times$ and $2.5\times$ slower than the classical baseline, respectively, with the noise model adding 16\% on top of ideal Qiskit. Because the classical head already provides near-perfect accuracy on this task, PennyLane's computational savings do not translate into a net advantage: the classical approach achieves higher accuracy at a moderate cost, making it the preferred choice for this dataset.

\subsubsection{Cats vs.\ Dogs}
\label{subsubsec:cats_dogs}

Table~\ref{tab:cats_detail} presents the per-backbone results on the Cats vs.\ Dogs dataset, the largest in our evaluation.

\begin{table}[htbp]
\centering
\caption{Cats vs.\ Dogs: Test accuracy and (training time in seconds) per backbone.}
\label{tab:cats_detail}
\resizebox{\textwidth}{!}{%
\begin{tabular}{lccccc}
\toprule
\textbf{Backbone} & \textbf{CC} & \textbf{PennyLane} & \textbf{PL-Noisy} & \textbf{Qiskit} & \textbf{Qiskit-Noisy} \\
\midrule
ResNet18        & 0.9160 (1157.5) & \textbf{0.9700} (832.8) & 0.9583 (1382.4) & 0.9700 (2282.8) & 0.9567 (2638.2) \\
MobileNetV2     & 0.9383 (979.9)  & \textbf{0.9800} (627.5) & 0.9717 (1038.6) & 0.9683 (2160.5) & 0.9550 (2497.1) \\
EfficientNet-B0 & 0.9500 (1358.6) & 0.9617 (740.2) & 0.9500 (1224.8) & \textbf{0.9650} (2226.5) & 0.9517 (2573.8) \\
RegNet-X-400MF  & 0.9200 (601.8)  & \textbf{0.9600} (460.4) & 0.9483 (762.1) & 0.9617 (2006.7) & 0.9450 (2319.7) \\
\bottomrule
\end{tabular}
}
\end{table}

Quantum-hybrid heads consistently outperform their classical counterparts on this dataset. PennyLane-Standard with MobileNetV2 achieves the highest accuracy (0.9800), surpassing the best classical result (EfficientNet-B0, 0.9500) by 3 percentage points. The gain is even more pronounced when comparing like-for-like backbones: PennyLane improves ResNet18 from 0.9160 to 0.9700, MobileNetV2 from 0.9383 to 0.9800, and RegNet-X-400MF from 0.9200 to 0.9600. Qiskit-Standard is equally competitive, matching PennyLane on ResNet18 (0.9700) and marginally surpassing it on EfficientNet-B0 (0.9650 vs.\ 0.9617) and RegNet-X-400MF (0.9617 vs.\ 0.9600). Regarding noise resilience, PL-Noisy exhibits a moderate degradation of 0.8--1.2\% compared with PennyLane ideal, while Qiskit-Noisy suffers a slightly larger drop of 1.3--1.7\% relative to Qiskit ideal. Despite this, all noisy variants still surpass or match the classical baselines except Qiskit-Noisy on EfficientNet-B0, which ties with the classical result (0.9517 vs.\ 0.9500). The overall pattern confirms that, on a medium-scale general-vision task, the 12-parameter quantum head adds measurable discriminative capacity over the frozen classical backbone alone.

Averaged across the four backbones, PennyLane-Standard is the fastest approach (665.2\,s), reducing training time by 35\% with respect to the classical baseline (1\,024.5\,s). This advantage stems from PennyLane's lightweight statevector execution, which processes a 16-sample batch in approximately 6\,ms and avoids the transpilation and scheduling overhead inherent to Qiskit. PL-Noisy increases the cost to 1\,102.0\,s (a 66\% overhead over PL ideal), explained by the shift from statevector to density-matrix simulation (\texttt{default.mixed}), which scales memory as $O(2^{2n})$ instead of $O(2^{n})$; nevertheless, it remains within 8\% of the classical baseline. Qiskit-Standard (2\,169.1\,s) and Qiskit-Noisy (2\,507.2\,s) are approximately $2.1\times$ and $2.4\times$ slower than the classical baseline, respectively, owing to circuit transpilation, finite-shot sampling ($N{=}1024$), and SPSA gradient estimation; the noise model adds a further 16\% on top of ideal Qiskit. PennyLane therefore offers the best accuracy-to-cost trade-off on this dataset, achieving the highest accuracy in the shortest time.

\subsubsection{Solar Dust}
\label{subsubsec:solar_dust}

Table~\ref{tab:solar_detail} presents the per-backbone results on Solar Dust, the smallest and most challenging dataset in our evaluation.

\begin{table}[htbp]
\centering
\caption{Solar Dust: Test accuracy and (training time in seconds) per backbone.}
\label{tab:solar_detail}
\resizebox{\textwidth}{!}{%
\begin{tabular}{lccccc}
\toprule
\textbf{Backbone} & \textbf{CC} & \textbf{PennyLane} & \textbf{PL-Noisy} & \textbf{Qiskit} & \textbf{Qiskit-Noisy} \\
\midrule
ResNet18        & 0.8250 (331.6) & 0.8333 (257.8) & 0.8167 (426.4) & \textbf{0.8833} (567.4) & 0.8667 (655.9) \\
MobileNetV2     & 0.8750 (300.7) & 0.8667 (238.2) & 0.8500 (394.1) & \textbf{0.9000} (537.3) & 0.8750 (621.1) \\
EfficientNet-B0 & 0.8583 (377.9) & 0.8667 (251.4) & 0.8417 (415.9) & \textbf{0.8583} (561.5) & 0.8333 (649.1) \\
RegNet-X-400MF  & \textbf{0.9250} (227.7) & 0.9167 (212.9) & 0.8917 (352.3) & 0.8750 (512.0) & 0.8500 (592.1) \\
\bottomrule
\end{tabular}
}
\end{table}

Solar Dust yields mixed results across frameworks. The classical RegNet-X-400MF achieves the overall best accuracy (0.9250), and PennyLane-Standard is within 0.8\% of it (0.9167). However, Qiskit-Standard provides the largest backbone-level improvements: ResNet18 rises from 0.8250 to 0.8833 and MobileNetV2 from 0.8750 to 0.90. PennyLane-Standard shows only marginal gains on the lighter backbones (ResNet18 +0.8\%, EfficientNet-B0 +0.8\%) and is slightly below the classical level on MobileNetV2 ($-$0.8\%). This dataset exhibits the highest noise sensitivity: PL-Noisy degrades by 1.7--2.5\% and Qiskit-Noisy by 1.7--2.5\% relative to their ideal counterparts, reflecting the limited statistical support that 300 samples provide for learning robust decision boundaries in the presence of stochastic quantum noise. The pattern suggests that Qiskit's shot-based execution, which implicitly exposes the optimiser to measurement uncertainty during training, builds greater robustness to noise on small, high-difficulty datasets.

PennyLane-Standard is again the fastest approach (240.1\,s on average), 22\% below the classical baseline (309.5\,s). PL-Noisy rises to 397.2\,s (a 65\% overhead over PL ideal), placing it 28\% above the classical cost, the highest relative overhead among all datasets, attributable to the density-matrix simulation scaling on a small batch count. Qiskit-Standard (544.6\,s) and Qiskit-Noisy (629.6\,s) are $1.8\times$ and $2.0\times$ slower than the classical baseline, respectively; the noise model adds 16\% over ideal Qiskit. Notably, the Qiskit-to-CC ratio is lower here than on the other three datasets ($1.8\times$ vs. $2.1$--$2.2\times$) because the small data volume reduces the number of circuit evaluations per epoch, partially amortising the fixed transpilation overhead.

\section{Conclusions}
\label{sec:conclusions}
This work presented a comparative evaluation of CQ hybrid transfer learning under realistic NISQ constraints, addressing key methodological challenges through unified hyperparameters, calibrated noise models derived from IBM quantum hardware, and heterogeneous image classification datasets. The study focused on assessing whether compact quantum heads can act as effective and efficient alternatives to classical classifiers when integrated with frozen, pretrained convolutional backbones.

Across the evaluated tasks, quantum-hybrid architectures achieved competitive and, in several cases, superior performance relative to classical baselines, particularly on datasets with limited sample sizes or high intra-class variability. Using noise parameters calibrated to IBM Heron r2 processor specifications ($T_1=250\,\mu$s, $T_2=150\,\mu$s), both PennyLane and Qiskit noisy variants showed a moderate degradation of 0.5--2.5\% in accuracy relative to ideal simulations, with smaller datasets such as Solar Dust showing higher sensitivity. The ideal PennyLane model reached peak accuracy of 97.44\% on \textit{Hymenoptera}. These results demonstrate that, with state-of-the-art quantum hardware characteristics, quantum-hybrid transfer learning models can maintain near-ideal performance even under realistic noise conditions.

Crucially, we extended our evaluation to real IBM Quantum hardware (\texttt{ibm\_torino}), demonstrating practical deployability of quantum transfer learning on current NISQ devices. Due to the substantial computational overhead of real QPU execution, including queue wait times, limited daily quotas, and the high cost of gradient estimation, these experiments were necessarily constrained to a single dataset (Hymenoptera) and reduced training duration (5 epochs). Despite these limitations, the Qiskit implementation with RegNet-X-400MF achieved 85.90\% test accuracy on real hardware, representing only a 5\% degradation relative to noisy simulation results and validating the predictive value of our calibrated noise models. This finding confirms that the performance trends observed in simulation translate meaningfully to actual quantum processors.

Crucially, we extended our evaluation to real IBM Quantum hardware (\texttt{ibm\_torino}), demonstrating practical deployability of quantum transfer learning on current NISQ devices. Due to the substantial computational overhead of real QPU execution, including queue wait times, limited daily quotas, and the high cost of gradient estimation, these experiments were necessarily constrained to the Hymenoptera dataset and a reduced training duration. Despite these limitations, the Qiskit implementation with MobileNetV2 achieved 94.87\% test accuracy on real hardware, only 2.6\% below its ideal simulation result and 0.6\% below the noisy simulation, validating the predictive value of our calibrated noise models. PennyLane-Real exhibited greater degradation (up to $-$8\% on MobileNetV2), attributable to the shift from the analytic parameter-shift rule to SPSA gradient estimation on the QPU. Notably, EfficientNet-B0 proved highly sensitive to hardware noise under both frameworks, dropping to 71.79\% (Qiskit) and 65.79\% (PennyLane), indicating that higher-dimensional feature spaces amplify the compounding effects of transpilation overhead and stochastic gradients on physical qubits. These findings confirm that the performance trends observed in simulation translate meaningfully to actual quantum processors, while also highlighting the critical role of backbone selection and gradient strategy in real-device deployments.

From a computational perspective, the study revealed a clear trade-off between simulation fidelity and training efficiency. PennyLane-based implementations consistently offered favorable accuracy–time trade-offs, outperforming both classical baselines and Qiskit-based hybrids in training speed. In contrast, Qiskit's shot-based sampling provided a more explicit treatment of measurement uncertainty. It exhibited stability in challenging datasets such as Solar Dust, albeit at the cost of substantially increased computational overhead. Taken together, these findings support the viability of QTL as a resource-efficient strategy, aligning with green AI objectives by reducing trainable parameters and limiting computational demands while maintaining competitive accuracy.

Despite these encouraging results, several limitations constrain the scope of the conclusions. Differences in noise modeling across frameworks complicate direct fidelity comparisons, and the limited number of runs prevents definitive statements regarding statistical robustness across random initializations. The real hardware experiments, while demonstrating practical viability, were limited to a single dataset and a reduced training protocol due to QPU access constraints. Moreover, the choice of gradient estimation method (parameter-shift vs. SPSA) significantly impacted real hardware performance, highlighting the need for hardware-aware optimization strategies.

Future work will extend this study along several complementary directions. Systematic evaluations across multiple random seeds are required to assess the stability of the observed performance trends. Adopting unified noise models across frameworks, including identical Pauli or Kraus channels, would enable more controlled cross-platform comparisons. In addition, broader architectural explorations that vary qubit counts, circuit depths, and entanglement patterns may help identify regimes in which hybrid quantum heads consistently provide benefits. Regarding real hardware deployment, extended training runs with error-mitigation techniques such as Zero-Noise Extrapolation, probabilistic error cancellation, or dynamical decoupling will be essential to close the gap between noisy simulation and real QPU performance. Finally, investigating hardware-efficient gradient methods that balance estimation accuracy with circuit evaluation cost remains a critical direction for practical quantum machine learning.

\section*{Data and code availability}
Codes and data are available at \url{https://github.com/Data-Science-Big-Data-Research-Lab/QTL}
\section*{Acknowledgments}
The authors would like to thank the Spanish Ministry of Science and Innovation for the support within the projects PID2023-146037OB-C21 and PID2023-146037OB-C22. We acknowledge Pablo de Olavide University for funding the Q-Resilience project. Finally, we acknowledge the use of IBM Quantum Credits for this work. The views expressed are those of the authors and do not reflect the official policy or position of IBM or the IBM Quantum team.

\bibliographystyle{amsplain}
\bibliography{refs}

@book{Goodfellow2016,
  author={Goodfellow, I. and Bengio, Y. and Courville, A.},
  title={Deep Learning},
  publisher={MIT Press},
  address={Cambridge, MA, USA},
  year={2016}
}

@article{Pan2010,
  author={Pan, S. J. and Yang, Q.},
  title={A Survey on Transfer Learning},
  journal={IEEE Transactions on Knowledge and Data Engineering},
  volume={22},
  number={10},
  pages={1345--1359},
  year={2010},
}

@inproceedings{Yosinski2014,
  author = {Yosinski, J. and Clune, J. and Bengio, Y. and Lipson, H.},
  title = {How Transferable are Features in Deep Neural Networks?},
  booktitle = {Advances in Neural Information Processing Systems 27},
  year = {2014},
  pages = {3320--3328},
  publisher = {Curran Associates, Inc.},
  address = {Montreal, QC, Canada}
}

@article{Zhuang2021,
  author={Zhuang, F. and Qi, Z. and Duan, K. and Xi, D. and Zhu, Y. and Zhu, H. and Zhou, J. and Huang, H. and He, Q.},
  title={A Comprehensive Survey on Transfer Learning},
  journal={Proceedings of the IEEE},
  volume={109},
  number={1},
  pages={43--76},
  year={2021},
}

@inproceedings{Deng2009,
  author={Deng, J. and Dong, W. and Socher, R. and Li, L.-J. and Li, K. and Li, F.-F.},
  title={ImageNet: A Large-Scale Hierarchical Image Database},
  booktitle={Proceedings of the IEEE Conference on Computer Vision and Pattern Recognition (CVPR)},
  pages={248--255},
  publisher={IEEE},
  address={Miami, FL, USA},
  year={2009},
}

@article{Russakovsky2015,
  author={Russakovsky, O. and Deng, J. and Su, H. and Krause, J. and Satheesh, S. and Ma, S. and Huang, Z. and Karpathy, A. and Khosla, A. and Bernstein, M. and Berg, A. C. and Li, F.-F.},
  title={ImageNet Large Scale Visual Recognition Challenge},
  journal={International Journal of Computer Vision},
  volume={115},
  number={3},
  pages={211--252},
  year={2015},
  publisher={Springer}
}

@article{Preskill2018,
  author={Preskill, J.},
  title={Quantum Computing in the {NISQ} Era and Beyond},
  journal={Quantum},
  volume={2},
  pages={79},
  year={2018}
}

@article{Schuld2015,
  author={Schuld, M. and Sinayskiy, I. and Petruccione, F.},
  title={An Introduction to Quantum Machine Learning},
  journal={Contemporary Physics},
  volume={56},
  number={2},
  pages={172--185},
  year={2015}
}

@article{Mitarai2018,
  author={Mitarai, K. and Negoro, M. and Kitagawa, M. and Fujii, K.},
  title={Quantum Circuit Learning},
  journal={Physical Review A},
  volume={98},
  number={3},
  pages={032309},
  year={2018}
}

@article{Havlicek2019,
  author={Havlíček, V. and Córcoles, A. D. and Temme, K. and Harrow, A. W. and Kandala, A. and Chow, J. M. and Gambetta, J. M.},
  title={Supervised Learning with Quantum-Enhanced Feature Spaces},
  journal={Nature},
  volume={567},
  pages={209--212},
  year={2019}
}

@article{Mari2020,
  author={Mari, A. and Bromley, T. R. and Izaac, J. and Schuld, M. and Killoran, N.},
  title={Transfer Learning in Hybrid Classical-Quantum Neural Networks},
  journal={Quantum},
  volume={4},
  pages={340},
  year={2020}
}

@misc{Farhi2018,
  author = {Farhi, E. and Neven, H.},
  title = {Classification with Quantum Neural Networks on Near Term Processors},
  howpublished = {arXiv preprint arXiv:1802.06002},
  year = {2018},
  note = {arXiv:1802.06002}
}

@inproceedings{He2016,
  author={He, K. and Zhang, X. and Ren, S. and Sun, J.},
  title={Deep Residual Learning for Image Recognition},
  booktitle={Proceedings of the IEEE Conference on Computer Vision and Pattern Recognition (CVPR)},
  pages={770--778},
  publisher={IEEE},
  address={Las Vegas, NV, USA},
  year={2016}
}

@inproceedings{Tan2019,
  author={Tan, M. and Le, Q. V.},
  title={EfficientNet: Rethinking Model Scaling for Convolutional Neural Networks},
  booktitle={Proceedings of the 36th International Conference on Machine Learning (ICML)},
  year={2019},
  pages={6105--6114},
  publisher={PMLR},
  address={Long Beach, CA, USA}
}

@misc{Kaggle2013,
  author={Cukierski, W.},
  title={Dogs vs. Cats Dataset},
  howpublished={Kaggle competition},
  year={2013},
  url={https://www.kaggle.com/c/dogs-vs-cats}
}

@misc{Nickparvar2021,
  author={Nickparvar, M.},
  title={Brain Tumor {MRI} Dataset},
  howpublished={Kaggle},
  year={2021},
  url={https://www.kaggle.com/datasets/masoudnickparvar/brain-tumor-mri-dataset}
}

@misc{TheDataSith2021,
  author={TheDataSith},
  title={Hymenoptera Dataset},
  howpublished={Kaggle},
  year={2021},
  url={https://www.kaggle.com/datasets/thedatasith/hymenoptera}
}

@misc{Garladinne2023,
  author={Garladinne-Hemanth, S.},
  title={Solar Panel Dust Detection Dataset},
  howpublished={Kaggle},
  year={2023},
  url={https://www.kaggle.com/datasets/hemanthsai7/solar-panel-dust-detection}
}

@article{Bali25,
  author={Bali, M. and Mishra, V. P. and Yenkikar, A. and Chikmurge, D.},
  title={{QuantumNet}: An Enhanced Diabetic Retinopathy Detection Model Using Classical Deep Learning--Quantum Transfer Learning},
  journal={MethodsX},
  volume={14},
  pages={103185},
  year={2025}
}

@article{Kim23,
  author = {Kim, J. and Huh, J. and Park, D. K.},
  title = {Classical-to-quantum convolutional neural network transfer learning},
  journal = {Neurocomputing},
  volume = {555},
  pages = {126643},
  year = {2023}
}

@article{Han23,
  author = {Han, K. and Huang, T. and Yin, L.},
  title = {Transfer learning accelerating complex parameters optimizations based on quantum-inspired parallel multi-layer Monte Carlo algorithm: Theory, application, implementation},
  journal = {Applied Soft Computing},
  volume = {134},
  pages = {109982},
  year = {2023}
}

@article{Mir21,
  author={Mir, A. and Yasin, U. and Khan, S. N. and Athar, A. and Jabeen, R. and Aslam, S.},
  title={Diabetic Retinopathy Detection Using Classical--Quantum Transfer Learning Approach and Probability Model},
  journal={Computers, Materials and Continua},
  volume={71},
  number={2},
  pages={3733--3754},
  year={2021}
}

@article{Wang24,
  author = {Wang, F. and Kebede, G. A. and Lo, S. and Woldegiorgis, B. H.},
  title = {An embedding layer-based quantum long short-term memory model with transfer learning for proton exchange membrane fuel stack remaining useful life prediction},
  journal = {Energy},
  volume = {308},
  pages = {133054},
  year = {2024}
}

@article{Otgonbaatar23,
  author={Otgonbaatar, S. and Schwarz, G. and Datcu, M. and Kranzlmüller, D.},
  title={Quantum Transfer Learning for Real-World, Small, and High-Dimensional Remotely Sensed Datasets},
  journal={IEEE Journal of Selected Topics in Applied Earth Observations and Remote Sensing},
  volume={16},
  pages={433--447},
  year={2023}
}

@inproceedings{Wang23,
  author = {Wang, R. and Du, J. and Gao, T.},
  title = {Quantum Transfer Learning Using the Large-Scale Unsupervised Pre-Trained Model WavLM-Large for Synthetic Speech Detection},
  booktitle = {Proceedings of the IEEE International Conference on Acoustics, Speech and Signal Processing (ICASSP)},
  year = {2023},
  pages = {1--5},
  publisher = {IEEE},
  address = {Rhodes Island, Greece}
}

@article{Buonaiuto24,
  author = {Buonaiuto, G. and Guarasci, R. and Minutolo, A. and De Pietro, G. and Esposito, M.},
  title = {Quantum transfer learning for acceptability judgements},
  journal = {Quantum Machine Intelligence},
  volume = {6},
  number = {1},
  pages = {13},
  year = {2024}
}

@article{Mogalapalli21,
  author={Mogalapalli, H. and Abburi, M. and Nithya, B. and Vamsi Bandreddi, S. K.},
  title={Classical--Quantum Transfer Learning for Image Classification},
  journal={SN Computer Science},
  volume={2},
  number={5},
  pages={1--13},
  year={2021}
}

@article{Azevedo22,
  author={Azevedo, V. and Silva, C. and Dutra, I.},
  title={Quantum Transfer Learning for Breast Cancer Detection},
  journal={Quantum Machine Intelligence},
  volume={4},
  pages={26},
  year={2022}
}

@article{RODRIGUEZ26,
  author={Rodríguez-Díaz, F. and Gutiérrez-Avilés, D. and Troncoso, A. and Martínez-Álvarez, F.},
  title={A Survey of Quantum Machine Learning: Foundations, Algorithms, Frameworks, Data and Applications},
  journal={ACM Computing Surveys},
  volume={58},
  number={4},
  pages={1--35},
  year={2025}
}

@inproceedings{Qi22,
  author={Qi, J. and Tejedor, J.},
  title={Classical-To-Quantum Transfer Learning for Spoken Command Recognition Based on Quantum Neural Networks},
  booktitle={Proceedings of the IEEE International Conference on Acoustics, Speech and Signal Processing (ICASSP)},
  pages={8627--8631},
  address={Singapore},
  publisher={IEEE},
  year={2022}
}

@inproceedings{Kumsetty22,
  author = {Kumsetty, N. V. and Bhat Nekkare, A. and S. Sowmya, K. and Kumar M., A.},
  title = {TrashBox: Trash Detection and Classification using Quantum Transfer Learning},
  booktitle = {Proceedings of the Conference of Open Innovations Association (FRUCT)},
  year = {2022},
  pages = {125--130},
  address = {Helsinki, Finland},
  publisher = {Open Innovations Association FRUCT}
}

@inproceedings{Koike22,
  author = {Koike-Akino, T. and Wang, P. and Wang, Y.},
  title = {Quantum Transfer Learning for Wi-Fi Sensing},
  booktitle = {Proceedings of the IEEE International Conference on Communications (ICC)},
  year = {2022},
  pages = {654--659},
  publisher = {IEEE},
  address = {Seoul, South Korea}
}

@article{Zen20,
  author={Zen, R. and My, L. and Tan, R. and Hébert, F. and Gattobigio, M. and Miniatura, C. and Poletti, D. and Bressan, S.},
  title={Transfer Learning for Scalability of Neural-Network Quantum States},
  journal={Physical Review E},
  volume={101},
  number={5},
  pages={053301},
  year={2020}
}

@article{Kati25,
  author = {Katı, B. E. and Küçüksille, E. U. and Sarıman, G.},
  title = {Enhancing Deepfake Detection Through Quantum Transfer Learning and Class-Attention Vision Transformer Architecture},
  journal = {Applied Sciences},
  volume = {15},
  number = {2},
  pages = {525},
  year = {2025}
}

@article{Yogaraj25,
  author = {Yogaraj, K. and Quanz, B. and Vikas, T. and Mondal, A. and Mondal, S.},
  title = {Post-variational classical quantum transfer learning for binary classification},
  journal = {Scientific Reports},
  volume = {15},
  number = {1},
  pages = {23682},
  year = {2025}
}

@article{Khatun25,
  author={Khatun, A. and Usman, M.},
  title={Quantum Transfer Learning with Adversarial Robustness for Classification of High-Resolution Image Datasets},
  journal={Advanced Quantum Technologies},
  volume={8},
  number={1},
  pages={2400268},
  year={2025}
}

@article{Vermeire21,
  author={Vermeire, F. H. and Green, W. H.},
  title={Transfer Learning for Solvation Free Energies: From Quantum Chemistry to Experiments},
  journal={Chemical Engineering Journal},
  volume={418},
  pages={129307},
  year={2021}
}

@inproceedings{Sandler2018mobilenetv2,
  author={Sandler, M. and Howard, A. and Zhu, M. and Zhmoginov, A. and Chen, L.-C.},
  title={{MobileNetV2}: Inverted Residuals and Linear Bottlenecks},
  booktitle={Proceedings of the IEEE Conference on Computer Vision and Pattern Recognition (CVPR)},
  year={2018},
  pages={4510--4520},
  publisher={IEEE},
  address={Salt Lake City, UT, USA}
}

@inproceedings{Radosavovic2020designing,
  author={Radosavovic, I. and Kosaraju, R. P. and Girshick, R. and He, K. and Dollár, P.},
  title={Designing Network Design Spaces},
  booktitle={Proceedings of the IEEE/CVF Conference on Computer Vision and Pattern Recognition (CVPR)},
  year={2020},
  pages={10428--10436},
  publisher={IEEE},
  address={Seattle, WA, USA}
}

@misc{Bergholm2018pennylane,
  author = {Bergholm, V. and Izaac, J. and Schuld, M. and Gogolin, C. and Ahmed, S. and Ajith, V. and Alam, M. S. and Alonso-Linaje, G. and AkashNarayanan, B. and Asadi, A. and Arrazola, J. M. and Azad, U. and Banning, S. and Blank, C. and Bromley, T. R. and Cordier, B. A. and Ceroni, J. and Delgado, A. and Di Matteo, O. and Dusko, A. and Garg, T. and Guala, D. and Hayes, A. and Hill, R. and Ijaz, A. and Isacsson, T. and Ittah, D. and Jahangiri, S. and Jain, P. and Jiang, E. and Khandelwal, A. and Kottmann, K. and Lang, R. A. and Lee, C. and Loke, T. and Lowe, A. and McKiernan, K. and Meyer, J. J. and Montañez-Barrera, J. A. and Moyard, R. and Niu, Z. and O'Riordan, L. J. and Oud, S. and Panigrahi, A. and Park, C.-Y. and Polatajko, D. and Quesada, N. and Roberts, C. and Sá, N. and Schoch, I. and Shi, B. and Shu, S. and Sim, S. and Singh, A. and Strandberg, I. and Soni, J. and Száva, A. and Thabet, S. and Vargas-Hernández, R. A. and Vincent, T. and Vitucci, N. and Weber, M. and Wierichs, D. and Wiersema, R. and Willmann, M. and Wong, V. and Zhang, S. and Killoran, N.},
  title = {PennyLane: Automatic differentiation of hybrid quantum-classical computations},
  howpublished = {arXiv preprint arXiv:1811.04968},
  year = {2018},
  note = {arXiv:1811.04968}
}

@misc{Qiskit,
  author={{Qiskit Contributors}},
  title={Qiskit: An Open-Source Framework for Quantum Computing},
  howpublished={\url{https://qiskit.org/}},
  year={2023}
}

@article{Sim2019,
  author={Sim, S. and Johnson, P. D. and Aspuru-Guzik, A.},
  title={Expressibility and Entangling Capability of Parameterized Quantum Circuits for Hybrid Quantum-Classical Algorithms},
  journal={Advanced Quantum Technologies},
  volume={2},
  number={12},
  pages={1900070},
  year={2019}
}

@misc{QiskitML,
  author={{Qiskit Machine Learning Contributors}},
  title={Qiskit Machine Learning},
  howpublished={\url{https://qiskit.org/ecosystem/machine-learning/}},
  year={2023}
}

@misc{QiskitAer,
  author={{Qiskit Aer Contributors}},
  title={Qiskit Aer: High Performance Simulator for Quantum Circuits},
  howpublished={\url{https://qiskit.org/ecosystem/aer/}},
  year={2023}
}

@article{SCHWARTZ20,
  title   = {Green {AI}},
  author  = {Schwartz, R. and Dodge, J. and Smith, N. A. and Etzioni, O.},
  journal = {Communications of the ACM},
  volume  = {63},
  number  = {12},
  pages   = {54-63},
  year    = {2020}
}

@article{LECUN15,
  title   = {Deep learning},
  author  = {Y. LeCun and Y. Bengio and G. Hinton},
  journal = {Nature},
  volume  = {521},
  number  = {12},
  pages   = {436-444},
  year    = {2015}
}

@article{BHARTI22,
  title   = {Noisy intermediate-scale quantum (NISQ) algorithms},
  author  = {Bharti, K. and Cervera-Lierta, A. and Kyaw, T. H. and others},
  journal = {Reviews of Modern Physics},
  volume  = {94},
  pages   = {015004},
  year    = {2022}
}

@article{DWIVEDI25,
  title={Efficient deep learning training: an energy-conscious adaptive framework},
  author={Dwivedi, Pulkit and Kajal, Mansi},
  journal={International Journal of Data Science and Analytics},
  volume={20},
  pages={4741-4755},
  year={2025}
}

@article{Spall1992,
  author={Spall, James C.},
  title={Multivariate Stochastic Approximation Using a Simultaneous Perturbation Gradient Approximation},
  journal={IEEE Transactions on Automatic Control},
  volume={37},
  number={3},
  pages={332--341},
  year={1992},
  publisher={IEEE}
}

@article{Guerreschi2020,
  author = {Guerreschi, Gian Giacomo and Smelyanskiy, Mikhail},
  title = {Practical optimization for hybrid quantum-classical algorithms},
  journal = {Scientific Reports},
  year = {2020},
  volume = {10},
  number = {1},
  pages = {4230},
}

@article{amaro2022filtering,
  author = {Amaro, D. and others},
  title = {Filtering variational quantum algorithms for combinatorial optimization},
  journal = {Quantum Science and Technology},
  year = {2022},
  volume = {7},
  number = {1},
  pages = {015013}
}

\end{document}